\documentclass{emulateapj}
\usepackage{amssymb,amsmath,graphicx,longtable,subfigure,wrapfig}
\usepackage{rotating}
\usepackage[colorlinks,linkcolor=blue,anchorcolor=green,citecolor=blue]{hyperref}

\begin{document}

\def\func#1{\mathop{\rm #1}\nolimits}
\def\unit#1{\mathord{\thinspace\rm #1}}

\title{A fallback accretion model for the unusual type II-P supernova
iPTF14hls}
\author{L. J. Wang\altaffilmark{1}, X. F. Wang\altaffilmark{2}, S. Q. Wang%
\altaffilmark{3,4,5}, Z. G. Dai\altaffilmark{3,4}, L. D. Liu%
\altaffilmark{3,4,6}, L. M. Song\altaffilmark{1}, L. M. Rui\altaffilmark{2},
Z. Cano\altaffilmark{7,8}, B. Li\altaffilmark{3,1}}

\begin{abstract}
The Intermediate Palomar Transient Factory reported the discovery of an
unusual type II-P supernova iPTF14hls. Instead of a $\sim $ 100-day plateau
as observed for ordinary type II-P supernovae, the light curve of iPTF14hls
has at least five distinct peaks, followed by a steep decline at $\sim 1000$%
\ days since discovery. Until 500 days since discovery, the effective
temperature of iPTF14hls is roughly constant at 5000-$6000\unit{K}$. In this
paper we propose that iPTF14hls is likely powered by intermittent fallback
accretion. It is found that the light curve of iPTF14hls can be well fit by
the usual $t^{-5/3}$\ accretion law until $\sim 1000$\ days post discovery
when the light curve transitions to a steep decline. To account for this
steep decline, we suggest a power-law density profile for the late accreted
material, rather than the constant profile as appropriated for the $t^{-5/3}$%
\ accretion law. Detailed modeling indicates that the total fallback mass is 
$\sim 0.2M_{\odot }$, with an ejecta mass $M_{\mathrm{ej}}\simeq 21M_{\odot
} $. We find the third peak of the light curve cannot be well fit by the
fallback model, indicating that there could be some extra rapid energy
injection. We suggest that this extra energy injection may be a result of a
magnetic outburst if the central object is a neutron star. These results
indicate that the progenitor of iPTF14hls could be a massive red supergiant.
\end{abstract}


\affil{\altaffilmark{1}Astroparticle Physics,
Institute of High Energy Physics,
Chinese Academy of Sciences, Beijing 100049, China; wanglingjun@ihep.ac.cn}

\affil{\altaffilmark{2}Physics Department and Tsinghua Center for Astrophysics,
 Tsinghua University, Beijing 100084, China}

\affil{\altaffilmark{3}School of Astronomy and Space Science, Nanjing
University, Nanjing 210093, China; dzg@nju.edu.cn}

\affil{\altaffilmark{4}Key Laboratory of
Modern Astronomy and Astrophysics (Nanjing University),
Ministry of Education, Nanjing 210093, China}

\affil{\altaffilmark{5}Department of Astronomy, University of California,
Berkeley, CA 94720-3411, USA}

\affil{\altaffilmark{6}Department of Physics and Astronomy, University of
Nevada, Las Vegas, NV 89154, USA}

\affil{\altaffilmark{7}Instituto de Astrof\'isica de Andaluc\'ia (IAA-CSIC),
Glorieta de la Astronom\'ia s/n, E-18008, Granada, Spain.}

\affil{\altaffilmark{8}Juan de la Cierva Fellow.}

\section{Introduction}

\label{sec:Intro}

Recently, the discovery of an unusual supernova (SN), iPTF14hls, was
reported by \cite{Arcavi17}. iPTF14hls, at a redshift of $z=0.0344$, was
first discovered in $R$ band on September 22, 2014 UT \citep{Arcavi17}.
Before its discovery, the position of iPTF14hls was not monitored for
approximately 100 days. At beginning, astronomers did not pay much attention
to iPTF14hls during its decline in brightness. Intense multiband
observations were deployed only when iPTF14hls began to rebrighten after
about 100 days since its discovery.

Although identified as a type II-P SN according to its spectroscopic
features \citep{Li15}, iPTF14hls is very unique among currently discovered
SNe. The light curve of iPTF14hls lasts for more than 1200 days %
\citep{Sollerman18} and has at least five distinct peaks, while an ordinary
type II-P SN has a 100-day plateau in brightness. The spectral evolution of
iPTF14hls is 10 times slower than typical SNe II-P \citep{Arcavi17}. The
photospheric velocities measured by Fe \textsc{ii }$\lambda $5169 stay at a
constant value of $4000\unit{km}\unit{s}^{-1}$.

\cite{Arcavi17} discussed several possible theoretical models, e.g.,
interaction between SN ejecta and circumstellar material 
\citep{Chevalier82,
Chevalier94, Chugai94, Chatzopoulos12, Moriya13, WangLiu2016, Wang18},
spin-down of a magnetar \citep{Kasen10, Woosley10}, fallback accretion onto
a black hole \citep{Michel88, Dexter13}, and suggested that the most likely
model may be fallback accretion. However, \cite{Dessart18} proposed that the
magnetar model can fit the light curve, while \cite{Soker18} explained
iPTF14hls as a common-envelope jets SN. \cite{Chugai18} and \citet{Woosley18}
discussed the models that might explain the light curve and spectral
features.

Here we suggest that the multiple peaks in the light curve of iPTF14hls
could be powered by intermittent fallback accretion of the SN ejecta. In a
successful SN, the material remaining bound could fallback and eventually
accrete onto the central object. Accretions onto compact objects (black
holes or neutron stars) are usually accompanied by powerful outflows 
\citep{Mirabel98,
Fender04}, which can carry away about 10\% of the gravitational binding
energy of the accreted material. Such powerful outflows can aid the
explosion of the SN, and on the other hand, a fraction of this energy would
be thermalized to power a bright light curve \citep{Dexter13}.

This paper is structured as follows. In Section \ref{sec:model} we describe
the fallback accretion process, while in Section \ref{sec:fit} the model and
fitting results are presented. Finally, we discuss and conclude our results
in Section \ref{sec:conc}.

\section{Fallback accretion}

\label{sec:model}

After the explosion of a core-collapse SN, a rebounce outward shock is
launched at the base of the central compact core, which further collapses
into a neutron star or black hole. This shock imparts a typical kinetic
energy of $\simeq 10^{51}\unit{erg}$ to the still-infalling material and
reverses it to move outward. The outward-moving material (ejecta) adjusts
itself quickly into a homologous expansion phase, that is, the expansion
velocity $v$ of a material element is proportional to its distance $r$ to
the central compact object. Although most of the ejecta becomes unbound to
the central compact object, a fraction of the ejecta with mass $M_{0}$ is
bound and finally falls back \citep{Colgate71} and accretes onto the central
compact object. Based on some arguments presented in Section \ref{sec:conc},
hereafter we assume that the remnant of iPTF14hls is a neutron star.

The material accreted at early times comes from the slowly moving inner
ejecta. Assuming a power-law density profile of the inner shell of the
progenitor star $\rho \left( r\right) =\rho _{0}\left( r/r_{0}\right)
^{\alpha -3}$, where $\rho _{0}$ is the density of the shell at radius $%
r_{0} $, the fallback accretion rate is 
\citep[for $0<\alpha<3$;][]{Quataert12,
Dexter13} 
\begin{equation}
\dot{M}=\frac{8\pi }{3-\alpha }\frac{\rho _{0}r_{0}^{3}}{t_{0}}\left( \frac{t%
}{t_{0}}\right) ^{\frac{3\left( \alpha -1\right) }{3-\alpha }}.
\end{equation}%
This accretion rate is usually rising because typically $1<\alpha <3$ for
inner shells. Here $t_{0}$ is defined as\footnote{%
Note that this definition of $t_{0}$ is different from that given in \cite%
{Dexter13} by an extra factor $\sqrt{\pi \alpha /8}$.}%
\begin{equation}
t_{0}=\left( \frac{\pi \alpha }{32G\rho _{0}}\right) ^{1/2}.
\end{equation}%
This accretion phase will transition to a long-term accretion phase when the
expansion velocity $v$ of the bound material is comparable to the escape
velocity $v_{\mathrm{esc}}$. In this case the material can reach a maximum
radius $r_{\max }=r_{0}\left( 1-v^{2}/v_{\mathrm{esc}}^{2}\right) ^{-1}$ %
\citep{Dexter13} and then falls back with a free-fall timescale $t_{\mathrm{%
ff}}$ \citep{Michel88}%
\begin{equation}
\frac{v_{\mathrm{esc}}^{2}-v^{2}}{v_{\mathrm{esc}}^{2}}=\left( \frac{t_{%
\func{col}}}{t_{\mathrm{ff}}}\right) ^{2/3},
\end{equation}%
where $t_{\func{col}}$ is the free-fall collapse time to form $M_{0}$ from
material at rest. Assuming a \emph{constant} density profile, the accretion
rate decays according to $t^{-5/3}$ \citep{Michel88}.

At very late phase, instead of a constant profile, the density may be a
steep power law, $\rho \left( r\right) =\rho _{0}\left( r/r_{0}\right)
^{\alpha -3}$ with $\alpha <0$, the enclosed mass is effectively constant,
and the accretion rate is \citep{Dexter13}%
\begin{equation}
\dot{M}=\frac{8\pi }{3}\frac{\rho _{0}r_{0}^{3}}{t_{1}}\left( \frac{t}{t_{1}}%
\right) ^{\left( 2\alpha -3\right) /3},
\end{equation}%
where%
\begin{equation}
t_{1}\equiv \pi \left( \frac{r_{0}^{3}}{8GM}\right) ^{1/2}.
\end{equation}%
Because the early rising phase in the light curve of iPTF14hls is missing,
we will model the light curve only by the $t^{-5/3}$ law and at very late
phase $t^{\left( 2\alpha -3\right) /3}$ law with some $\alpha <0$.

Assuming a spherical accretion, \cite{Chevalier89} and \cite{Houck91}
studied the structure of the accretion flow that may operate in the famous
SN 1987A. To power an SN like iPTF14hls by accretion, the accretion rate
(see Section \ref{sec:fit}) should be high (in the range $10^{-4}\lesssim 
\dot{M}_{\mathrm{ac}}\lesssim 10^{4}M_{\odot }\unit{yr}^{-1}$) and the
gravitational accretion energy is carried away by neutrinos produced near
the neutron star \citep{Chevalier89, Houck91}. However, this does not mean
that the photons in the accretion flow cannot heat the ejecta.

To determine whether the radiation advected with the accretion flow is able
to diffuse out, the trapping radius 
\citep{Katz77, Begelman78,
Flammang82, Blondin86}%
\begin{equation}
r_{\mathrm{tr}}=\frac{\dot{M}_{\mathrm{ac}}\kappa }{4\pi c}=5.5\times
10^{13}\left( \frac{\dot{M}_{\mathrm{ac}}}{M_{\odot }\unit{yr}^{-1}}\right)
\left( \frac{\kappa }{0.33\unit{cm}^{2}\unit{g}^{-1}}\right) \unit{cm}
\end{equation}%
is defined at which the inwardly advected radiation flux balances the
outward diffusion flux. Photons outside this radius can diffuse out and heat
the ejecta, while the photons inside this radius are trapped. Because
iPTF14hls is hydrogen-rich, here we take the electron Thomson scattering
opacity $\kappa =0.33\unit{cm}^{2}\unit{g}^{-1}$ %
\citep[e.g.,][]{Moriya11,Chatzopoulos12}, which is suitable for fully
ionized material with solar metallicity.

The inner regions of the accretion flow achieve supersonic free fall %
\citep{Chevalier89}, which, upon reaching the neutron star surface,
generates a strong shock moving outward. The energy is mainly stored inside
but close to the shock radius. The shock radius $r_{s}$\ is determined by
neutrino cooling efficiency. Photons inside $r_{s}$\ act as potential
heating source of the SN. Whether the photons inside $r_{s}$\ can diffuse
out depends on if the condition $r_{s}>r_{\mathrm{tr}}$\ is satisfied. Here
we simply assume that this condition is satisfied and leave the
justification in Section \ref{sec:conc}.

At the accretion rate mentioned above, the accretion is super-Eddington.
During accretion the infalling material is compressed and becomes hot and
geometrically thick because of the inability of the advected photons to
escape from the accretion flow. As a result, the accretion is accompanied by
powerful outflow \citep{Narayan94, Blandford99, Igumenshchev00, McKinney12},
as verified by the observation of ultra-relativistic outflow from a neutron
star accreting gas from a companion \citep{Fender04}. Usually the accretion
rate is assumed to be a power-law in radius $\dot{M}\left( r\right) =\dot{M}%
_{\mathrm{fb}}\left( r/r_{\mathrm{fb}}\right) ^{s}$ 
\citep[e.g.,][]{Kohri05,
Dexter13}, where $\dot{M}_{\mathrm{fb}}$ is the mass accretion rate at the
fallback radius $r_{\mathrm{fb}}$, and $0<s<1$. It should be stressed that $%
\dot{M}_{\mathrm{fb}}$ is not the mass accretion rate onto the compact
object because a large fraction of the accretion flow is channeled as an
outflow. The net accretion rate $\dot{M}_{\mathrm{ac}}$ onto the neutron
star is usually only $\sim 1\%$ of $\dot{M}_{\mathrm{fb}}$, namely $\dot{M}_{%
\mathrm{ac}}=\xi \dot{M}_{\mathrm{fb}}$\ with $\xi \simeq 0.01$. About 10\%
of the accreted matter is converted as radiation energy. Consequently, the
accretion energy rate is $\dot{E}_{w}=\epsilon \dot{M}_{\mathrm{fb}}c^{2}$
with $\epsilon \simeq 10^{-3}$ \citep{Dexter13}.

Because of the existence of powerful outflows, the accretion cannot be
strictly spherical. The aspherical accretion may be induced by the spiral
modes of the standing accretion shock instability 
\citep[SASI;][]{Burrows95,
Janka96, Blondin03, Marek09, Fernandez10} or the convection in the
pre-collapse envelope \citep{Gilkis14}. Recently, the jet-feedback mechanism %
\citep{Gilkis16} based on SASI is suggested to carry out the accretion
energy. In this scenario the energy may be carried out by jets accreted near
the equatorial plane, as indicated by observations \citep{Fender04}.

The fallback mass is a function of the compactness of the progenitor stars
and explosion energy \citep{Chevalier89, Zhang08}. For loose progenitors
with typical explosion energies $\sim 10^{51}\unit{erg}$, like red
supergiants (RSGs), the fallback mass is usually small, $\lesssim
0.1M_{\odot }$. However, for more compact progenitors, e.g., blue
supergiants, the H/He interface triggers the formation of a strong reverse
shock, which decelerates the ejecta and enhances the fallback mass
significantly. For weak explosions, most of the mass may fall back %
\citep{Moriya10}. The metallicity of the progenitor stars influences the
mass loss history before explosion and therefore is another factor that
impacts the fallback mass. As a result, fallback accretion influences the
final mass of the central compact objects. For population III (zero
metallicity) stars above $25M_{\odot }$ and explosion energies less than $%
1.5\times 10^{51}\unit{erg}$, the central compact objects are more likely
black holes \citep{Zhang08} because of large amount of fallback. For
population I (solar metallicity) stars, black hole production is much less
frequent because of large scale mass loss before explosion.

\section{The model and fitting results}

\label{sec:fit}

The accretion outflows not only heat the SN ejecta, but also accelerate the
ejecta. We use the method outlined in \cite{WangWang16} to calculate the
light curve and the evolution of the photospheric velocities. In this model
the photospheric radius is at the position outside of which the optical
depth is equal to $2/3$ \citep{WangWang16}. The acceleration of the ejecta
by the energy injection has been taken into account by this model, which\
assumes a homologous expansion of the SN ejecta, with a homogeneous density
distribution. The energy injection from the energy sources, which may be a
spinning-down magnetar, $^{56}$Ni cascade decay, or fallback accretion, will
be trapped by the ejecta. The trapped energy undergoes adiabatic expansion,
which accelerates the ejecta according to the following equation %
\citep{WangWang16}%
\begin{equation}
\frac{dE_{K}}{dt}=L_{\mathrm{inp}}-L_{e},
\end{equation}%
where $L_{\mathrm{inp}}$\ is the power trapped by the ejecta, $L_{e}$\ is
the SN luminosity, and $E_{K}$\ is the kinetic energy of the SN. The
expansion velocity $v_{\mathrm{sc}}$\ (which is approximately equal to the
observed photospheric velocity for massive ejecta at early epoch) is
calculated according to $E_{K}=3M_{\mathrm{ej}}v_{\mathrm{sc}}^{2}/10$\ %
\citep{Arnett82}, where $M_{\mathrm{ej}}$\ is the ejecta mass. A part of the
trapped energy diffuses out of the ejecta, resulting in the multiband
optical emission of the SN.

To account for the multiple peaks in the light curve of iPTF14hls, we
propose that the accretion is episodic. Such episodes are not rare in
astrophysics. For example, episodic accretion may be caused by instabilities
of disks around protostars \citep{Sakurai16,
Kuffmeier18}, or by knotty jets in young protostellar disks %
\citep{Vorobyov18}.

For fallback accretion, the energy input is%
\begin{equation}
L_{\mathrm{inp}}=\dot{E}_{w}=\epsilon \dot{M}_{\mathrm{fb}}\left( t\right)
c^{2},  \label{eq:fallback-power}
\end{equation}%
where $\dot{M}_{\mathrm{fb}}\left( t\right) $ takes the expression%
\begin{equation}
\dot{M}_{\mathrm{fb}}\left( t\right) =\dot{M}_{i}\left( t/t_{i}\right)
^{-5/3}
\end{equation}%
during the constant density accretion phase and%
\begin{equation}
\dot{M}_{\mathrm{fb}}\left( t\right) =\dot{M}_{i}\left( t/t_{i}\right)
^{-\left( 2\alpha -3\right) /3}
\end{equation}%
for the final power-law density accretion. Here $\dot{M}_{i}$ is the mass
fallback rate at time $t_{i}$ when the $i$th fallback episode begins. To
calculate the light curve of an SN powered by fallback accretion, the energy
input given by Equation $\left( \ref{eq:fallback-power}\right) $\ takes the
place of the magnetar spinning-down power in the case of a magnetar-powered
SN. In both the magnetar-powered case and the accretion-powered case, the
energy is assumed to be deposited at the center of the SN ejecta. The
photospheric emission is a result of photon diffusion.

In this work we use the bolometric luminosity data of iPTF14hls provided by 
\cite{Sollerman18} who extended the observation to more than 1200 days since
discovery. We neglect the possible contribution of $^{56}$Ni and $^{56}$Co
to the light curve of iPTF14hls. The SN explosion would have surely
synthesized some amount of $^{56}$Ni. However, because of the finite
lifetimes of $^{56}$Ni ($8.8\unit{days}$) and $^{56}$Co ($111.3\unit{days}$%
), such contribution is only limited to the first $\sim 100\unit{days}$
since explosion, which were largely missed by the observation.

\begin{figure}[tbph]
\centering\includegraphics[width=0.5\textwidth,angle=0]{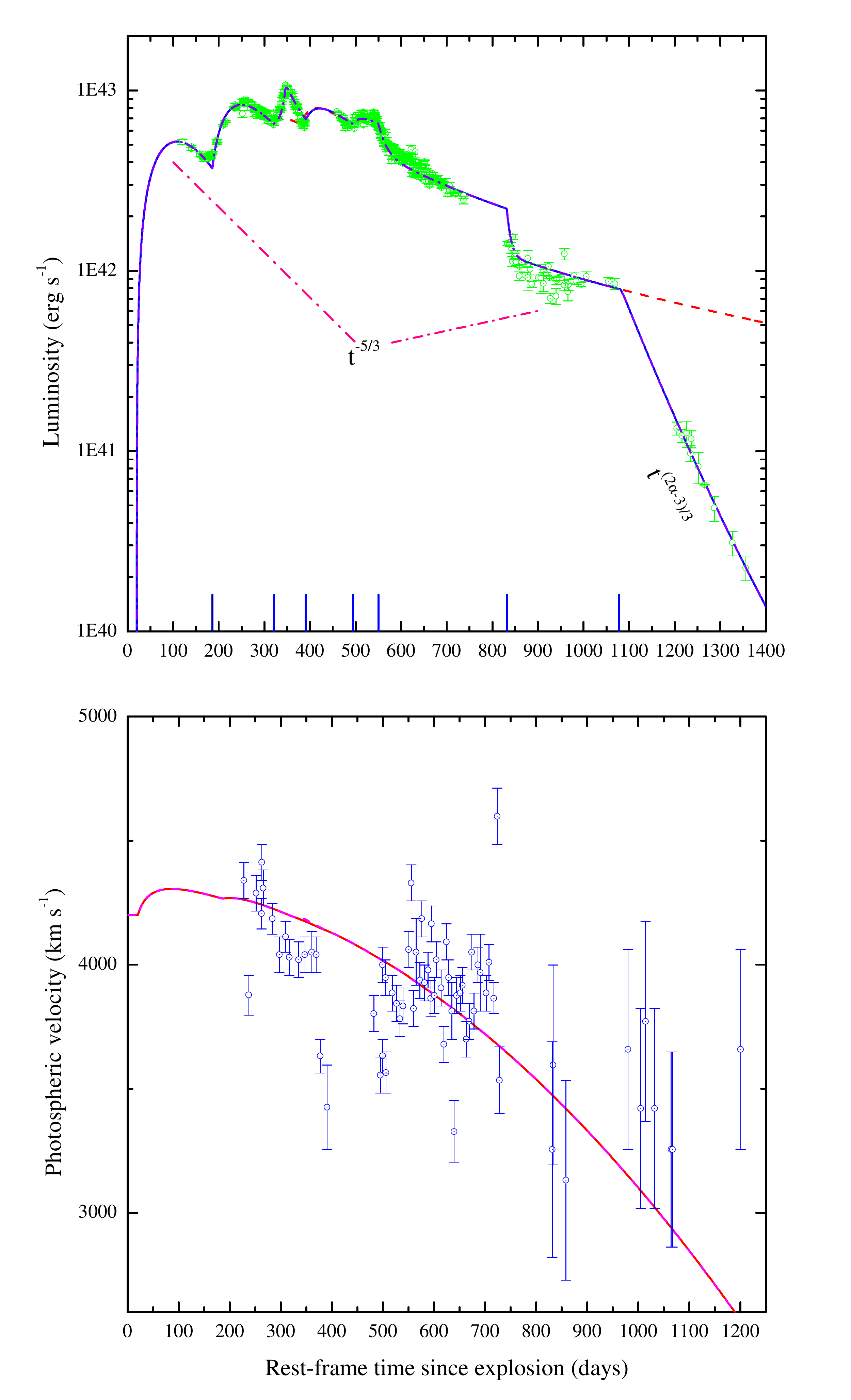}
\caption{Light curve (top panel) and photospheric velocity (bottom panel) of
iPTF14hls reproduced by the fallback accretion model (solid lines). The
dashed lines are the modeling results including the energy injection from an
outburst. The red dashed line in the upper panel assumes a $t^{-5/3}$
accretion rate, which fails to fit the data after $\sim 1100$ days since
explosion. This is why a late-time steep decline has been introduced here,
as depicted by the solid blue line.}
\label{fig:lc-v}
\end{figure}

The fitting results (solid lines), including the light curve and
photospheric velocity evolution, are shown in Figure \ref{fig:lc-v}, with
the 19 fitting parameters listed in Table \ref{tbl:para}. It can be found
that the fallback rates listed in Table \ref{tbl:para} are similar to that
given by \cite{Moriya18}, who interpret OGLE-2014-SN-073 as a fallback
accretion powered type II supernova. To give a decent fit to the light
curve, eight episodes are needed. In Figure \ref{fig:lc-v} we mark $t_{i}$
as vertical blue ticks. It is found that the first seven accretion episodes
can be fit by the $t^{-5/3}$\ law, whereas the last episode ($t_{8}$\ and $%
\dot{M}_{8}$) can only be fit by a steep decay with a density power-law
index $\alpha \simeq -22$. With this $\alpha $, the late-time light curve
decay index is $\sim -15.6$, slightly steeper than that measured by \cite%
{Sollerman18}, who gave a decay index $-13.5$.

\begin{table*}[tbph]
\caption{Best-fitting parameters.}
\label{tbl:para}
\begin{center}
\begin{tabular}{ccccccccccccccccccc}
\hline\hline
$M_{\mathrm{ej}}$ & $v_{\mathrm{sc}0}$ & $\dot{M}_{1}$ & $t_{1}$ & $\dot{M}%
_{2}$ & $t_{2}$ & $\dot{M}_{3}$ & $t_{3}$ & $\dot{M}_{4}$ & $t_{4}$ & $\dot{M%
}_{5}$ & $t_{5}$ & $\dot{M}_{6}$ & $t_{6}$ & $\dot{M}_{7}$ & $t_{7}$ & $\dot{%
M}_{8}$ & $t_{8}$ & $\alpha $ \\ \hline
21 & 4200 & 4.9 & 20 & 0.75 & 186 & 0.43 & 321 & 0.5 & 380 & 0.4 & 494 & 0.24
& 550 & 0.065 & 832 & 0.043 & 1078 & $-22$ \\ \hline
\end{tabular}%
\end{center}
\par
\textbf{Notes.} $M_{\mathrm{ej}}$ and $v_{\mathrm{sc}0}$ are in units of $%
M_{\odot }$ and $\unit{km}\unit{s}^{-1}$, respectively. The accretion rates $%
\dot{M}_{i}$ at fallback radius are in units of $10^{-8}M_{\odot }\unit{s}%
^{-1}$, while $t_{i}$ are in units of $\unit{days}$ since SN explosion. In
this fit we fixed $\kappa =0.33\unit{cm}^{2}\unit{g}^{-1}$. Because of the
lack of observational data between the third and fourth peaks, $t_{4}$
cannot be accurately constrained, so is $t_{1}$ because of the missing of
observational data around the first peak. The first seven accretion episodes
can be fit by the $t^{-5/3}$ law, whereas the last episode ($t_{8}$ and $%
\dot{M}_{8}$) can only be fit by a steep decay with a density power-law
index $\alpha $.
\end{table*}

The explosion date is $\sim 120\unit{days}$ before the first observational
data point. In Table \ref{tbl:para} $v_{\mathrm{sc}0}$\ is the initial
expansion velocity of the surface of the ejecta. Assuming homologous
expansion of the ejecta, the initial explosion energy of this SN is $3M_{%
\mathrm{ej}}v_{\mathrm{sc}0}^{2}/10=2.2\times 10^{51}\unit{erg}$. This
energy can be attributed to neutrino-driven mechanism, which may drive an
explosion up to energy $\sim 2.5\times 10^{51}\unit{erg}$\ 
\citep{Janka16,
Bollig17}.

Figure \ref{fig:lc-v} shows that the fallback accretion model gives a
reasonably good fit to both the light curve and velocity evolution of
iPTF14hls. However, the third peak in the light curve cannot be fitted. Such
a steep peak require a very rapid energy release rate. This may suggest some
activity in the central compact object. At such late times ($\sim 300\unit{%
days}$) since explosion, the energy may be released by a magnetic outburst %
\citep{Gavriil02, Rea09, Rea12} if the central object is a neutron star.
Indeed, stellar evolution model predicts that a single star with initial
masses between $\sim 8$ and $25M_{\odot }$ will explode as an SN II-P,
leaving behind a neutron star remnant \citep{Heger03}.

To quantify the outburst power, we assume a power-law injection%
\begin{equation}
L_{\mathrm{rise}}=L_{\mathrm{pk}}\left( \frac{t-t_{\mathrm{start}}}{t_{%
\mathrm{rise}}}\right) ^{n},  \label{eq:L-rise}
\end{equation}%
followed by a rapid shutoff of the outburst%
\begin{equation}
L_{\mathrm{fall}}=L_{\mathrm{pk}}\left( \frac{t_{\mathrm{shutoff}}-t}{t_{%
\mathrm{fall}}}\right) .
\end{equation}%
Here $t_{\mathrm{start}}$ and $t_{\mathrm{shutoff}}$ are the times at which
the outburst begins and ends, respectively; $t_{\mathrm{rise}}$ and $t_{%
\mathrm{fall}}$ are the durations for the rise and fall of the outburst,
respectively. Obviously, $t_{\mathrm{shutoff}}-t_{\mathrm{start}}=t_{\mathrm{%
rise}}+t_{\mathrm{fall}}$. We set $n=3$ in Equation $\left( \ref{eq:L-rise}%
\right) $ during the fitting. This power exponent does not result from the
fitting constraints but was appropriately selected. To give a good fit to
the light curve, we found $2\lesssim n\lesssim 5$. The 5 fitting parameters
for the outburst are listed in Table \ref{tbl:burst-para}, with the
resulting light curve depicted in Figure \ref{fig:lc-v} as dashed lines.
Here we choose another set of values for $t_{4}$\ and $\dot{M}_{4}$\ (see
Section \ref{sec:conc} for some discussion). It can be seen that the model
including a magnetic outburst can fit the light curve very closely.

\begin{table}[tbph]
\caption{Fitting parameters for the outburst.}
\label{tbl:burst-para}
\begin{center}
\begin{tabular}{ccccc}
\hline\hline
$t_{\mathrm{start}}$ & $t_{\mathrm{rise}}$ & $t_{\mathrm{fall}}$ & $L_{%
\mathrm{pk}}$ & $n$ \\ \hline
$\left( \unit{days}\right) $ & $\left( \unit{days}\right) $ & $\left( \unit{%
days}\right) $ & $\left( \unit{erg}\unit{s}^{-1}\right) $ &  \\ \hline
$318$ & $29.2$ & $0.8$ & $1.7\times 10^{43}$ & $3$ \\ \hline
\end{tabular}%
\end{center}
\end{table}

From Table \ref{tbl:burst-para} we see that the outburst lasted for $\sim 30%
\unit{days}$, and released $1.1\times 10^{49}\unit{erg}$ in total. This is
in accordance with observations, which show that some X-ray pulsars may
experience sporadic giant X-ray outbursts lasting weeks to years followed by
a long-term quiescence \citep{Gavriil02, Kaspi03, Rea12, Cusumano16}.

\begin{figure}[tbph]
\includegraphics[width=0.5\textwidth,angle=0]{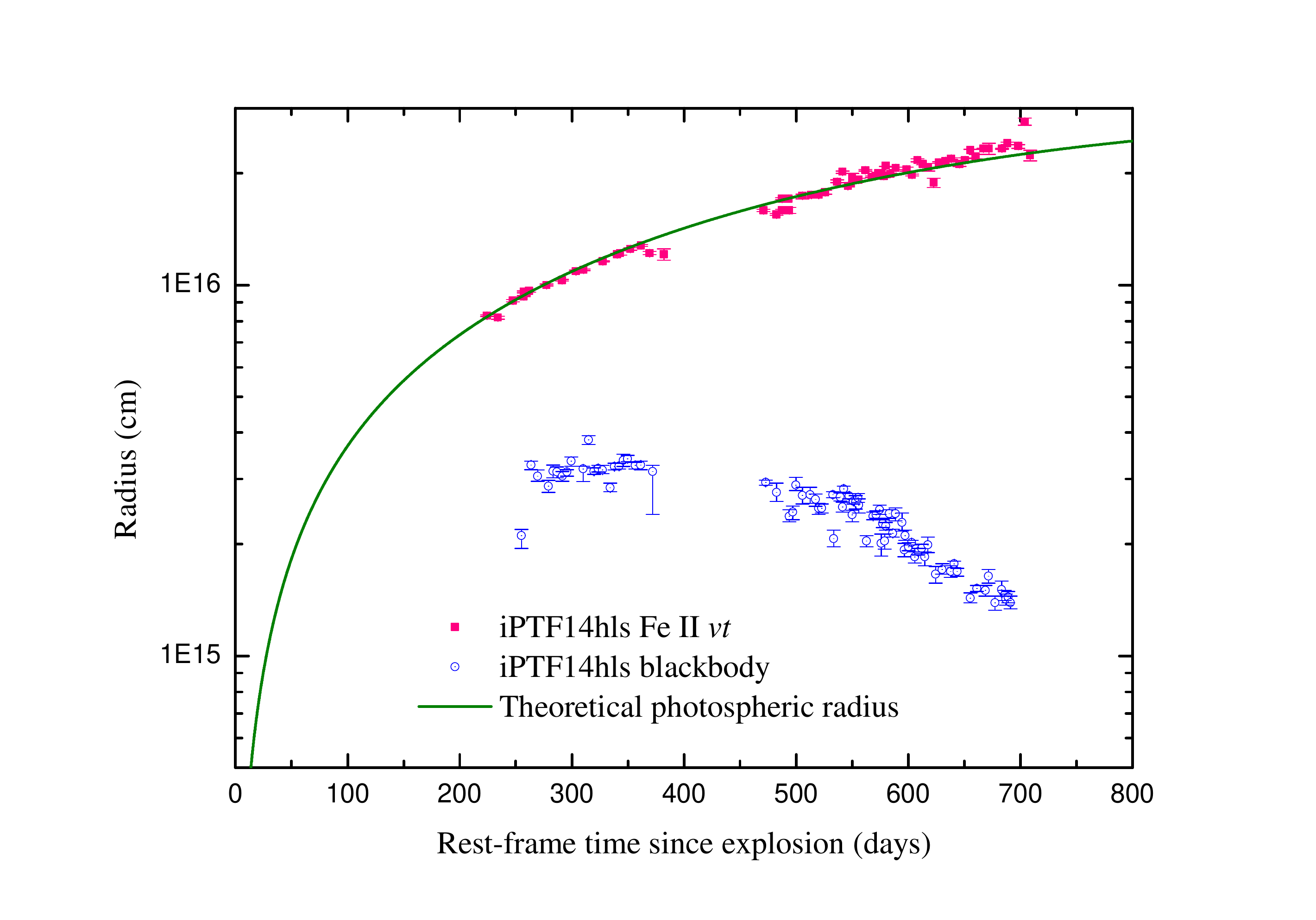} \centering
\caption{Fit (solid green line) to the photospheric radius of iPTF14hls
estimated (1) using blackbody fits to the broad-band $BVgi$ photometry
(blue) and (2) using the expansion velocities of Fe \textsc{ii }$\protect%
\lambda 5169$ times the elapsed rest-frame time ($vt$) since explosion
(pink). The data are taken from \protect\cite{Arcavi17}, but rescaled to
times since explosion, rather than since discovery.}
\label{fig:ph-radius}
\end{figure}

One intriguing feature of iPTF14hls is that the radius derived by the Fe 
\textsc{ii }$\ \lambda 5169$\ expansion velocity times the elapsed
rest-frame time is not equal to the radius determined by blackbody fits (see
Figure \ref{fig:ph-radius}). Spectrum measurements of Fe \textsc{ii }\ $%
\lambda 5169$\ indicate that iron expands at roughly a constant velocity $%
\sim 4000\unit{km}\unit{s}^{-1}$, from which the so-called line-forming
radius can be derived (pink points in Figure \ref{fig:ph-radius}). This
radius is, however, much larger than the blackbody-determined radius (blue
circles in Figure \ref{fig:ph-radius}). As can be seen from Figure \ref%
{fig:ph-radius}, the photospheric radius predicted by this model closely
follows the line-forming radius. We discuss this feature in Section \ref%
{sec:conc}.

\section{Discussion and conclusions}

\label{sec:conc}

To date, many SNe were found to be double-peaked 
\citep{Arnett89,
Richmond94, Mazzali08, Nicholl15, Nicholl16}, in which case the first
short-lived peak has been attributed to shock cooling 
\citep{Piro15,
Vreeswijk17, WangSQCano17}, although sometimes the cooling peak could merge
with the second main peak \citep{Wang18}. The first peak in the light curve
of iPTF14hls is unlikely the result of shock cooling because that would
require a very massive and extended envelope.

Because of the lack of observational data between the third and fourth peaks
of the light curve, and also the missing of observational data around the
first peak, $t_{1}$ and $t_{4}$ in Table \ref{tbl:para} cannot be accurately
constrained. As a demonstration of this uncertainty, in Figure \ref{fig:lc-v}
we choose different $t_{4}$ for the solid and dashed curves. However, as can
be seen from Figure \ref{fig:lc-v}, an earlier $t_{4}$, as depicted by the
dashed curve, is preferred because of the light-curve decline rate between $%
t_{4}$ and $t_{5}$. For the dashed line (the curve including the energy
injection of a magnetic outburst), we choose $t_{4}=391\unit{days}$ and $%
\dot{M}_{4}=0.49\times 10^{-8}M_{\odot }\unit{s}^{-1}$.

The most likely progenitor of iPTF14hls is a red supergiant since
observations have demonstrated that the progenitors of several type II-P SNe
are RSGs (e.g., \citealt{Smartt09,Davies17,VanDyk17,Huang18}). This is
consistent with the fact that the total fallback mass is about $\sim
0.2M_{\odot }$, as expected for a RSG progenitor 
\citep{Chevalier89,
Dexter13}. The ejecta mass, $M_{\mathrm{ej}}\simeq 21M_{\odot }$, is also
consistent with a RSG \citep{Davies18}, though at the high end of the
distribution of the SNe II-P ejecta masses.

The remnant of a RSG explosion is believed to be a neutron star. This is
consistent with the study of remnant mass distribution of massive star
explosions \citep{Zhang08}. iPTF14hls occurred on the outskirts of a
low-mass star-forming galaxy, indicating low metallicity \citep{Arcavi17}.
For population III star explosions with progenitor mass larger than $%
25M_{\odot }$\ and explosion energies less than $1.5\times 10^{51}\unit{erg}$%
, black holes are more frequent outcome. The progenitor mass of iPTF14hls, $%
22M_{\odot }$\ (the sum of $M_{\mathrm{ej}}$\ and remnant mass, which is
assumed to be $\sim 1M_{\odot }$), combined with its explosion energy $%
2.2\times 10^{51}\unit{erg}$, indicates that the remnant of iPTF14hls is
more likely a neutron star. On the other hand, despite the possible low
metallicity of iPTF14hls, \cite{Arcavi17} estimated a metallicity of $%
0.5Z_{\odot }$, which is more compatible with a population I star explosion.
In this case the production of a black hole remnant is highly suppressed %
\citep{Zhang08}.

The production of a neutron star remnant is also partially supported by the
need to fit the third peak of the light curve by a magnetic outburst.
Inspection of Table \ref{tbl:para} shows that the mass fallback rates $\dot{M%
}_{i}$ decrease monotonically, as expected. However, $\dot{M}_{1}$ is much
larger than $\dot{M}_{2}$. This is in sharp contrast to the mass fallback
rates that follow. $\dot{M}_{1}$ could be reduced if there is some
contribution from the energy injection of the neutron star. We have
neglected the contribution of $^{56}$Ni and $^{56}$Co because their
contribution is short-lived, but the contribution of a neutron star (or
magnetar) could be long-lived 
\citep{Kasen10, Woosley10, Inserra13,
Nicholl14, Metzger15, WangWang15, WangYu17, Dai16, LiuWang17}.

To examine the properties (dipole magnetic field $B_{p}$, and initial spin
period $P_{0}$) of the neutron star, we set (somewhat arbitrarily) $\dot{M}%
_{1}=1.0\times 10^{-8}M_{\odot }\unit{s}^{-1}$ and found $P_{0}\simeq 8\unit{%
ms}$, $B_{p}\simeq 5\times 10^{14}\unit{G}$. With these parameters, the
first peak can be closely fitted while the remaining peaks are affected
negligibly by the contribution of the magnetar. We will not show the fitting
results by this model because the resulting light curve closely follows the
curves presented in Figure \ref{fig:lc-v}. It is found that $P_{0}$ cannot
be too large, say $P_{0}\gtrsim 10\unit{ms}$, because in that case the
magnetar would contribute too much at late times so that the late-time light
curve deviates from the $t^{-5/3}$ law. We note that the above constraints
on $P_{0}$ and $B_{p}$ should not be taken seriously because they are
degenerated with $\dot{M}_{1}$.

As we said, the third peak cannot be explained by fallback accretion, and
magnetic activity is therefore proposed as its energy source. For the
fallback rate at the third peak, $\dot{M}_{\mathrm{fb}}\simeq 0.4\times
10^{-8}M_{\odot }\unit{s}^{-1}$, the corresponding accretion rate $\dot{M}_{%
\mathrm{ac}}=\xi \dot{M}_{\mathrm{fb}}\simeq 1\times 10^{-3}M_{\odot }\unit{%
yr}^{-1}$\ indicates a trapping radius $r_{\mathrm{tr}}\simeq 7\times 10^{10}%
\unit{cm}$. For the energy in the magnetic activity to diffuse out, the
dissipation radius $r_{\mathrm{act}}$\ of the magnetic energy should be
larger than $r_{\mathrm{tr}}$, namely $r_{\mathrm{act}}>r_{\mathrm{tr}}$. To
estimate $r_{\mathrm{act}}$, let us first assume a spherical accretion. The
balance of stellar wind pressure $L/4\pi r^{2}c$\ with the ram pressure $%
\rho v^{2}=\left( \dot{M}_{\mathrm{ac}}/4\pi \right) \left( 2GM/r^{5}\right)
^{1/2}$\ of the infalling material gives a radius%
\begin{equation}
r_{b}=2GM\left( \frac{\dot{M}_{\mathrm{ac}}c}{L}\right) ^{2}\simeq 1.3\times
10^{12}\unit{cm},  \label{eq:r_b}
\end{equation}%
where the typical value of pulsar luminosity $L=1\times 10^{41}\unit{erg}%
\unit{s}^{-1}$\ at the time of the third peak, the neutron star mass $%
M=1.4M_{\odot }$, accretion rate $\dot{M}_{\mathrm{ac}}=\xi \dot{M}_{\mathrm{%
fb}}=0.01\times 10^{-8}M_{\odot }\unit{s}^{-1}$\ have been substituted. This
balance is unstable. During the accretion phase, mater falls within $r_{b}$.
During the magnetic outburst phase, the central neutron star inflates a
bubble, known as pulsar wind nebula (PWN), whose radius $r_{\mathrm{act}}$\
is much larger than $r_{b}$, which is also larger than $r_{\mathrm{tr}}$. As
a result, the magnetic energy stored within the PWN can diffuse out of the
accretion flow.

During the normal accretion phase the balance between the magnetic pressure
and the accretion ram pressure cannot be maintained and the material falls
well within $r_{b}$. The magnetic outburst is usually triggered by some
instability of the PWN. The energy released by the spinning-down pulsar does
not lose immediately as radiation. It is estimated that about half of the
energy lost by Crab ($\sim 1.8\times 10^{49}\unit{erg}$) is still resident
within the synchrotron nebula \citep{Hester08}. This energy is very close to
the energy assumed here to power the third peak of iPTF14hls. The magnetic
activity may interplay with and even quench the accretion. If the accretion
is quenched by the magnetic activity, then the third peak is purely powered
by the magnetic activity. This argument also applies to the first peak where
a magnetar spin-down was proposed to contribute most part of the SN
luminosity.

For a rapidly spinning magnetar, outside of the light cylinder, $%
R_{c}=cP/(2\pi )=3.8\times 10^{7}\left( P/8\unit{ms}\right) \unit{cm}$, the
magnetic field lines of the magnetar cannot corotate with the magnetar %
\citep{Shapiro83}, and therefore the field lines in the PWN wind tightly to
form a spindle nebula \citep[see Figure 3 of ][]{Hester08}, whose toroidal
field is amplified significantly and much stronger than its poloidal field.
The lower bound of $r_{\mathrm{act}}$\ given by Equation $\left( \ref{eq:r_b}%
\right) $\ yields an upper limit of the strength of the magnetic field
within the PWN, $B_{\mathrm{PWN}}<\left( 8\pi E_{\mathrm{burst}}/V\right)
^{1/2}=5\times 10^{6}\unit{G}$, where $V=4\pi r_{b}^{3}/3$\ is the lower
limit of the volume of the PWN, and $E_{\mathrm{burst}}\sim 1.1\times 10^{49}%
\unit{erg}$\ is the magnetic outburst energy.

The total magnetic outburst energy, $E_{\mathrm{burst}}\sim 1.1\times 10^{49}%
\unit{erg}$, should be accumulated during the first $\sim 320\unit{days}$\
before the third peak. This requires an average energy injection rate $%
4\times 10^{41}\unit{erg}\unit{s}^{-1}$\ during this period.\footnote{%
The true energy injection rate should be somewhat higher because a part of
the injected energy will leak out of the PWN.} We found that with the
magnetar parameters ($P_{0}\simeq 8\unit{ms}$, $B_{p}\simeq 5\times 10^{14}%
\unit{G}$) listed above to explain the first peak of iPTF14hls, it is just
right to give such an average energy injection rate. At day $1400$\ since
explosion, the magnetar's energy injection rate declines to $1\times 10^{40}%
\unit{erg}\unit{s}^{-1}$, which is lower than but comparable to the observed
luminosity of iPTF14hls. After $1400\unit{days}$\ since explosion, the
magnetar's energy injection rate dominates over the accretion energy
injection rate. Therefore the suggestion of magnetic outburst scenario for
the third peak can be falsified if future observation does not reveal a
flattening of the luminosity of iPTF14hls after $1400\unit{days}$.

As mentioned in Section \ref{sec:fit}, the initial explosion energy ($%
2.2\times 10^{51}\unit{erg}$) of this SN can be attributed to neutrino
heating. However, for an SN that is powered by fallback accretion, the
explosion energy may also be partially provided by accretion, especially the
recently proposed jet-feedback mechanism \citep{Gilkis16, Soker16,
Soker17}.

We mentioned in Section \ref{sec:Intro} that for spherical accretion, the
photons behind the accretion shock can diffuse out and heat the ejecta if
the condition $r_{s}>r_{\mathrm{tr}}$\ is fulfilled. Assuming a power-law
neutrino cooling function, the shock position takes the approximate form %
\citep{Houck91}%
\begin{equation}
r_{s}=1.6\times 10^{8}\left( \frac{\dot{M}_{\mathrm{ac}}}{M_{\odot }\unit{yr}%
^{-1}}\right) ^{-2/5}\unit{cm}.
\end{equation}%
With the peak accretion rate $\dot{M}_{\mathrm{ac}}\simeq 10^{-3}M_{\odot }%
\unit{yr}^{-1}$, the above equation gives $r_{s}=2.5\times 10^{9}\unit{cm}$,
which is at its face value smaller than the trapping radius $r_{\mathrm{tr}%
}\simeq 7\times 10^{10}\unit{cm}$. However, the above estimate of $r_{s}$\
should be treated as a lower limit because of the uncertainties in neutrino
cooling function and relativistic corrections \citep{Houck91}. It is
actually found that the condition $r_{s}>r_{\mathrm{tr}}$\ is satisfied when 
$\dot{M}_{\mathrm{ac}}\lesssim 10^{-3}M_{\odot }\unit{yr}^{-1}$\ %
\citep{Houck91}, which is the case for the accretion episodes listed in
Table \ref{tbl:para}, except for the first accretion episode because for the
fallback rate $\dot{M}_{\mathrm{fb}}\simeq 0.7\times 10^{-8}M_{\odot }\unit{s%
}^{-1}$\ (see Table \ref{tbl:para}), the accretion rate is $\dot{M}_{\mathrm{%
ac}}=\xi \dot{M}_{\mathrm{fb}}\simeq 2\times 10^{-3}M_{\odot }\unit{yr}^{-1}$%
. For the scenario proposed in this paper to be valid, a magnetar energy
input is necessary for the first accretion episode. This also strengthens
the hypothesis of the formation of a magnetar in this SN explosion.

With the condition $r_{s}>r_{\mathrm{tr}}\simeq 7\times 10^{10}\unit{cm}$,
the shock radius $r_{s}$\ is about four orders of magnitude larger than the
Schwarzschild radius and it seems unlikely to convert 10\% of the
gravitational binding energy of the accreted material into radiation energy.
However, the exact accretion process is that the accretion flow reaches at
the neutron star surface, where shock is formed and energy is advected along
with the outmoving shock and carried far away from the neutron star surface.
The shock eventually stops at $r_{s}$\ because of efficient neutrino
cooling. During this process, a significant fraction (approximately 10\%) of
the gravitational binding energy is converted into shock energy. Note that
the fallback energy conversion factor $\epsilon \simeq 10^{-3}$\ consists of
two factors: the ratio of accretion rate to the fallback rate ($\xi =\dot{M}%
_{\mathrm{ac}}/\dot{M}_{\mathrm{fb}}\simeq 0.01$) and the conversion
efficiency (10\%) of the gravitational binding energy.

To account for the late steep decline of the light curve, we suggest that
accreted material has a power-law density profile with $\alpha =-22$\ at
late time. If we adopt the light curve decay index $-13.5$, as measured by 
\cite{Sollerman18}, we found $\alpha =-18$. This density profile is very
steep and may be formed by the interaction between the bound and unbound
material. Future numerical simulations are encouraged to test this
hypothesis.

We propose that the third, brightest peak is mainly powered by a magnetic
outburst. Such outburst is usually accompanied by X-ray emission, which is
however not detected \citep{Arcavi17}. The nondetection of X-ray emission
can be understood by considering the optical depth of the ejecta in the
X-ray band%
\begin{eqnarray}
\tau _{X} &=&\frac{3\kappa _{X}M_{\mathrm{ej}}}{4\pi v_{\mathrm{sc}}^{2}t^{2}%
}  \notag \\
&=&20\left( \frac{M_{\mathrm{ej}}}{21M_{\odot }}\frac{\kappa _{X}}{0.33\unit{%
cm}^{2}\unit{g}^{-1}}\right) \left( \frac{v_{\mathrm{sc}}}{4300\unit{km}%
\unit{s}^{-1}}\right) ^{-2}\left( \frac{t}{350\unit{days}}\right) ^{-2},
\label{eq:x-depth}
\end{eqnarray}%
where the values of X-ray opacity $\kappa _{X}$, SN expansion velocity $v_{%
\mathrm{sc}}$, and the time since explosion $t$ have been substituted. Here $%
v_{\mathrm{sc}}$ is slightly larger than the initial expansion velocity $v_{%
\mathrm{sc}0}$ because of the energy injection. We see that at the time the
third peak was observed, the ejecta are still opaque to X-rays. In the above
estimate, $\kappa _{X}$ is taken to be the same as $\kappa $, that is, the
electron Thomson scattering opacity. This should be a lower limit to the
true X-ray opacity because other heavier elements could make a significant
contribution to $\kappa _{X}$.

Despite the nondetection of X-ray emission, the detection of $\gamma $-ray
emission, temporally and positionally consistent with iPTF14hls, in the
energy band between 0.2 and $500\unit{GeV}$\ was report by \cite{Yuan18}.
The $\gamma $-ray source appears $\sim 300\unit{days}$\ after the first
optical detection of iPTF14hls and is still detectable up to $\sim 850\unit{%
days}$. Translated to the time since SN explosion in our model, the $\gamma $%
-ray source appears $\sim 420\unit{days}$\ to $\sim 970\unit{days}$.
According to Equation $\left( \ref{eq:x-depth}\right) $, assuming a lower
limit to $\gamma $-ray opacity $\kappa _{\gamma }=0.33\unit{cm}^{2}\unit{g}%
^{-1}$,\footnote{%
At late stage, the $\gamma $-ray photons come from $^{56}$Co decay with
typical energy $\sim 1\unit{MeV}$. At this energy, the atomic scattering
opacity \citep{Kotera13} of a type II SN is approximately equal to the
electron Thomson scattering opacity.} the SN ejecta are still opaque to $%
\gamma $-ray emission at time $t\sim 970\unit{days}$. Therefore, in our
model the $\gamma $-ray emission cannot come from the deep interior of the
SN ejecta.

This $\gamma $-ray emission may alternatively result from the interaction
between the ejecta and circumstellar medium (CSM) or produced by a blazar
because there is a blazar candidate within the error circle of the $\gamma $%
-ray source \citep{Yuan18}. Late-time observation of iPTF14hls revealed
narrow H$\alpha $\ emission \citep{Andrews18}, which may be evidence for
circumstellar interaction where unshocked circumstellar material is ionized
by the shock emission and recombines. However, such evidence for interaction
only appears at 3 years after the first optical detection of iPTF14hls. The
interaction origin of the $\gamma $-ray emission is also in tension with the
aforementioned nondetection of X-ray and radio emission. \cite{Sollerman18}
argue that the narrow H$\alpha $\ emission may come from H {\scriptsize II}
region which is located just at the SN position. Because the $\gamma $-ray
association with iPTF14hls is only tentative, we consider it more likely
that the $\gamma $-ray emission is produced by the blazar.

Observations indicate that the photospheric radius of iPTF14hls is quite
different from the line-forming region. \cite{Arcavi17} estimate the latter
at position of $vt$, where $v$ is the SN expansion velocity. Although the
photospheric radius of an SN recedes as the SN expands and inner material is
observed, the large ejecta mass, as inferred from light curve modeling,
implies that the photospheric recession should be negligible during the
first two years since its discovery. The discrepancy of these two radii
might be linked to the existence of persistent Balmer series P Cygni lines
observed in the spectra of iPTF14hls \citep{Arcavi17}. The presence of
P-Cygni profiles betrays the existence of a stellar wind, as observed in
Wolf-Rayet stars \citep{Willis82} and luminous blue variables %
\citep{Israelian99}. We suggest that this wind is far above the photosphere
and is responsible for the spectral lines.

The rarity of iPTF14hls among SNe II-P may be understood because of its
extreme ejecta mass. This large ejecta mass may also account for the reason
why so much mass falls back so as to give a multi-peaked light curve.

In summary, iPTF14hls can be explained by the episodic fallback accretion
model.\footnote{%
Alternatively, the CSM interaction is also a plausible model. %
\citet{LiuWang18} propose a multiple ejecta-CSM interaction model and
employed it to model multi-peak SNe iPTF15esb and iPTF13dcc.} The fitting
parameters suggest a RSG as the progenitor. Although the central object
cannot be identified, the rapid third peak and other considerations might
indicate the formation of a neutron star that experienced a magnetic
outburst lasting for $\sim 30\unit{days}$ with a total burst energy $%
1.1\times 10^{49}\unit{erg}$.

\begin{acknowledgements}
We thank Iair Arcavi and Jesper Sollerman for providing us the observational data.
We also thank the anonymous referee for helpful comments.
This work is supported by the National Program on Key Research and Development
Project of China (Grant Nos. 2016YFA0400801 and 2017YFA0402600), National
Basic Research Program of China (\textquotedblleft 973" Program, Grant
No. 2014CB845800) and the National Natural Science Foundation of China (Grant
Nos. 11573014, 11533033, 11673006). X. Wang is supported by the National Natural Science 
Foundation of China (NSFC grants 11325313 and 11633002), and the National 
Program on Key Research and Development Project (grant no. 2016YFA0400803).
S.Q.W. and L.D.L. are also supported by China Scholarship Program to conduct research 
at U.C. Berkeley and UNLV, respectively.
\end{acknowledgements}

\end{document}